\documentclass{article}%
\usepackage{amsmath}
\usepackage{amsfonts}
\usepackage{amssymb}
\usepackage{graphicx}%
\providecommand{\U}[1]{\protect\rule{.1in}{.1in}}

\begin{document}

\begin{center}
\bigskip\textbf{LIGHT BENDING IN THE GALACTIC HALO BY RINDLER-ISHAK METHOD }

\bigskip

\textbf{Amrita Bhattacharya,}$^{1,2,a}$\textbf{ Ruslan Isaev,}$^{3,b}$\textbf{
Massimo Scalia,}$^{2,c}$ \textbf{Carlo Cattani,}$^{2,4,d}$\textbf{ and Kamal
K. Nandi}$^{1,3,5,e}$

\bigskip

$^{1}$Department of Mathematics, University of North Bengal, Siliguri 734 013, India

$^{2}$Dipartimento di Matematica, Istituto \textquotedblleft
G.Castelnuovo\textquotedblright, Universit\`{a} La Sapienza, P.le Aldo Moro,
2, Rome, Italy

$^{3}$Joint Research Laboratory, Bashkir State Pedagogical University, Ufa
450000, Russia

$^{4}$DiFarma, Universit\`{a} di Salerno, Via Ponte Don Melillo, 84084
Fisciano, Salerno, Italy

$^{5}$Department of Theoretical Physics, Sterlitamak State Pedagogical
Academy, Sterlitamak 453103, Russia

\bigskip

$^{a}$Email: amrita\_852003@yahoo.co.in

$^{b}$Email: subfear@gmail.com

$^{c}$Email: Massimo.Scalia@uniroma1.it

$^{d}$Email: ccattani@unisa.it

$^{e}$E-mail: kamalnandi1952@yahoo.co.in
\end{center}

\bigskip

PACS number(s): 04.50.1h, 04.20.Cv

\begin{center}
\bigskip

\textbf{Abstract}
\end{center}

After the work of Rindler and Ishak, it is now well established that the
bending of light is influenced by the cosmological constant $\Lambda$
appearing in the Schwarzschild-de Sitter spacetime. We show that their method,
when applied to the galactic halo gravity parametrized by a constant $\gamma$,
yields exactly the same $\gamma-$ correction to Schwarzschild bending as
obtained by standard methods. Different cases are analyzed, which include some
corrections to the special cases considered in the original paper by Rindler
and Ishak.

\begin{center}
-------------------------------------------------------------
\end{center}

\textbf{I. Introduction}

It has long been believed that light bending in the Schwarzschild-de Sitter
(SdS) spacetime is uninfluenced by the cosmological constant $\Lambda$
appearing in the metric. The reason is that $\Lambda$ cancels out of the
second order null geodesic equation $-$ naturally, the light trajectory too
does not contain it. On the other hand, $\Lambda$ appears in the first order
differential equation, only its further differentiation removes the constant
$\Lambda$ from the second order differential equation. Obviously, for the sake
of consistency, the perturbative solution of second order equation must
satisfy the first order equation as well, which would then yield a relation
among the involved constants, one of which is the impact parameter $b$. This
would in turm imply that the removed $\Lambda$ will reappear in the geometric
light trajectory, hence in the light bending as well. We shall address this
question in more detail elsewhere. Here we shall limit ourselves only to the
perturbative solution of the second order equation expressed customarily in
terms of closest approach distance.\footnote{Following Weinberg [16], we shall
take in the light orbit the closest approach distance $R$ in preference to the
impact parameter $b$. This then implies that we can carry on with the second
order differential equation with its perturbative solution involving $R$. It
should be mentioned that Weinberg's integral too eventually involves only $R$
(not $b$) in which $\Lambda$ disappears (see [13]). The method has been
applied only to the asymptotically flat spacetimes, which is \textit{not} the
case here. Thus, a new method, such as that of Rindler and Ishak, seems more
appropriate, which we follow here.}

Recently, a new method has been proposed by Rindler and Ishak [1] that
combines the standard perturbative solution with an invariant geometric
definition of the bending angle. The method reveals that the Schwarzschild
bending caused by a lens (say, a galactic cluster of mass $M$) is decreased by
repulsive bending due to $\Lambda$. This work has instantly generated a lot of
interest among the gravity physics community (see, e.g., [3,4,5,6,7,8,9,10]).
It can be fairly said that the original work by Rindler and Ishak (as well as
its extension [2] to the Einstein-Strauss vacuole model) has established
beyond doubt that there \textit{is} a $\Lambda-$ dependent effect on light
bending contrary to previous belief, although its interpretation in the
cosmological set up is still open to further discussion and research (see the
recent review [11]). In view of this new wisdom, it would be interesting to
apply the Rindler-Ishak method to the domain of galactic halo gravity. For
this purpose, an excellent example seems to be the Mannheim-Kazanas-de Sitter
(MKdS) solution of Weyl gravity with a conformal parameter $\gamma$,
interpreted as a player in the halo gravity. The Weyl theory is based on the
15-parameter group of conformal invariance and it attempts to resolve the dark
matter/dark energy problem without hypothesizing them. It should be noted
however that the MKdS solution has its own merits and demerits, but their
discussion is beyond the scope of this paper. Our interest lies only in
finding if the Rindler-Ishak method can reproduce the first order effect of
$\gamma$ on light bending already known in the literature. We show that it
indeed does, which can be treated as yet another success of the method.

The purpose of the present paper is to implement in a more general spacetime
the Rindler-Ishak method to one higher order in $M$ than considered
originally, and calculate the effect of $\gamma$ on the bending of light rays.
It turns out that the method delivers the \textit{exact} first order $\gamma-$
term in addition to revealing new interplays among the constants $\Lambda$,
$M$ and $\gamma$. Different cases are discussed and some needed corrections in
[1] are pointed out.

The contents are organized as follows: In Sec.II, we derive the geodesic
equation in the MKdS solution. In Sec.III, we work out the bending of light
rays following the Rindler-Ishak method. We discuss specific cases in Sec.IV
while Sec.V summarizes the paper. There are three appendices.

\textbf{II. Geodesic equation }

An interesting solution of the Weyl gravity field equations is the MKdS metric
given by [12,13] (in units $8\pi G=c_{0}=1$):%
\begin{equation}
d\tau^{2}=B(r)dt^{2}-\frac{1}{B(r)}dr^{2}-r^{2}(d\theta^{2}+\sin^{2}\theta
d\varphi^{2}),\text{ \ }B(r)=1-\frac{2M}{r}+\gamma r-kr^{2},\text{\ }%
\end{equation}
where $M$ is the luminous central mass, $k$ and $\gamma$ are constants. The
numerical value of $k=\Lambda/3=0.43\times10^{-56}$cm$^{-2}$, and $\gamma$ is
of the order of inverse Hubble length. However, we caution that there is a
reported ambiguity both in the magnitude and sign of $\gamma$. By analyzing
the flat rotation curve data, Mannheim and Kazanas [12] fix it to be positive,
$\gamma\approx+10^{-28}$ cm$^{-1}$, while Pireaux [14] argues for
$\gamma\approx-10^{-33}$ cm$^{-1}$. Edery and Paranjape [13] conclude from the
time delay data in the halo that $\gamma\simeq-7\times10^{-28}$ cm$^{-1}$. We
emphasize that we are free to adduce any sign to $\gamma$ in the sequel but
for definiteness, we choose $\gamma>0$ in the present work. Such choice is
neither mandatory nor essential for the present work.

Denoting $u=1/r$, we derive the following path equation for a test particle of
mass $m_{0}$ on the equatorial plane $\theta=\pi/2$ as follows:%
\begin{equation}
\frac{d^{2}u}{d\varphi^{2}}=-u+3Mu^{2}-\frac{\gamma}{2}+\frac{M}{h^{2}}%
+\frac{1}{2h^{2}u^{2}}\left(  \gamma-\frac{2k}{u}\right)  ,
\end{equation}
where $h=\frac{J}{m_{0}}$, the angular momentum per unit test mass. For
photon, $m_{0}=0\Rightarrow h\rightarrow\infty$ and one ends up with the null
geodesic equation without $k$ making its appearance:%
\begin{equation}
\frac{d^{2}u}{d\varphi^{2}}=-u+3Mu^{2}-\frac{\gamma}{2}.
\end{equation}
Here we find that a cancellation of $k$ similar to that in the SdS case
($\gamma=0$) occurs despite the presence of a nonzero $\gamma$ in the metric.
Exactly like in the SdS case, one would now tend to believe that the bending
of light in the MKdS case would not be the influenced by $k$ to any order but
this is not the case.

\textbf{III. Rindler-Ishak method }

We shall follow the three principles adopted by Rindler and Ishak [1]:
\textbf{(i)} The method is originally applied to the SdS metric to show that,
despite the non-appearance of $k$ in the null geodesic, its effect still
appears in light bending. Hence to preserve the essence of the method, we
shall retain $k\neq0$ except in special cases. \textbf{(ii)} With $M\neq0$,
$k\neq0$ in the metric, the limit $r\rightarrow\infty$ makes no sense. The
only intrinsically characterized $r$ value \textit{replacing} $r\rightarrow
\infty$ is the one at $\varphi=0$. The measurable quantities are the various
$\psi$ angles that the photon orbit makes with successive coordinate planes
$\varphi=$ const. We shall qualitatively verify the results for $\varphi\neq0$
as well, say, at $\varphi=\pi/4$. \textbf{(iii)} We shall use the perturbative
solution for $\frac{1}{r}$ up to order $M^{2}$ and the resulting deflection
angle $\psi$.

Let us develop the basic equations now. Although the MKdS metric is more
general than the SdS metric, we shall show that the influence of $\gamma$
appears in the light bending together with that of $k$ including terms mixing
up the two. The light path equation (3) in the zeroth order is%
\begin{equation}
\frac{d^{2}u_{0}}{d\varphi^{2}}+u_{0}=\frac{\gamma}{2}%
\end{equation}
and its exact solution is%
\begin{equation}
u_{0}=\frac{\cos\varphi}{R}-\frac{\gamma}{2}%
\end{equation}
where $R$ is the distance of closest approach to the origin (where the lens
is). This solution is to be used as the zeroth order approximation. Following
the usual method of small perturbations [15], we want to derive the solution
as%
\begin{equation}
u=u_{0}+u_{1}+u_{2}%
\end{equation}
where $u_{1}$ and $u_{2}$ respectively satisfy%
\begin{equation}
\frac{d^{2}u_{1}}{d\varphi^{2}}+u_{1}=3Mu_{0}^{2}%
\end{equation}%
\begin{equation}
\frac{d^{2}u_{2}}{d\varphi^{2}}+u_{2}=6Mu_{0}u_{1}.
\end{equation}
The exact solutions are%
\begin{equation}
u_{1}=\frac{M}{4R^{2}}\left[  6+3R^{2}\gamma^{2}-6R\gamma\cos\varphi
-2\cos2\varphi-6R\gamma\varphi\sin\varphi\right]
\end{equation}%
\begin{align}
u_{2}  &  =\frac{3M^{2}}{16R^{3}}[\left\{  10+3R^{2}\gamma^{2}\left(
5-2\varphi^{2}\right)  \right\}  \cos\varphi-12R\gamma(4+R^{2}\gamma
^{2})\nonumber\\
&  +\cos3\varphi+20\varphi\sin\varphi+16R\gamma\cos2\varphi+30R^{2}\gamma
^{2}\varphi\sin\varphi+8R\gamma\varphi\sin2\varphi].
\end{align}
Formally changing $\varphi\rightarrow\frac{\pi}{2}-\varphi$, the final
solution up to second order in $M$ can be written as%
\begin{align}
u  &  \equiv\frac{1}{r}=\frac{\sin\varphi}{R}-\frac{\gamma}{2}\nonumber\\
&  +\frac{M}{4R^{2}}\left[  6+3R^{2}\gamma^{2}-3R\gamma(\pi-2\varphi
)\cos\varphi+2\cos2\varphi-6R\gamma\sin\varphi\right] \nonumber\\
&  -\frac{3M^{2}}{32R^{3}}[96R\gamma+24R^{3}\gamma^{3}-10(2+3R^{2}\gamma
^{2})(\pi-2\varphi)\cos\varphi+32R\gamma\cos2\varphi\nonumber\\
&  -20\sin\varphi-30R^{2}\gamma^{2}\sin\varphi+3\pi^{2}R^{2}\gamma^{2}%
\sin\varphi-12\pi R^{2}\gamma^{2}\varphi\sin\varphi\nonumber\\
&  +12R^{2}\gamma^{2}\varphi^{2}\sin\varphi-8\pi R\gamma\sin2\varphi
+16R\gamma\varphi\sin2\varphi+2\sin3\varphi].
\end{align}
This perturbative expansion holds only for small $u$ or large $r$. Thus we are
considering parameters $M$, $R$ and $\gamma$ such that $\frac{M}{R}<1$ and
$\gamma R<1$. When $\gamma=0$, it may be verified that one recovers the
equation for light trajectory up to $M^{2}$ order in the Schwarzschild metric.
The following clarification be noted: In the expression for $u$ in Ref.[2], a
trivial term $10\sin\varphi$ is deleted while another trivial term $10\pi
\cos\varphi$ is retained. From the standpoint of generality, we keep our exact
orbit equation as it is, but show that the presence or absence of such trivial
terms do \textit{not} lead to any difference whatsoever in the final result
(See Appendix C).

The method of Rindler and Ishak is based on the invariant formula for the
cosine of the angle $\psi$ between two coordinate directions $d$ and $\delta$
such that%
\begin{equation}
\cos\psi=\frac{g_{ij}d^{i}\delta^{j}}{(g_{ij}d^{i}d^{j})^{1/2}(g_{ij}%
\delta^{i}\delta^{j})^{1/2}}.
\end{equation}
Differentiating Eq.(11) with respect to $\varphi$, and denoting $\frac
{dr}{d\varphi}\equiv A(r,\varphi)$, we get%
\begin{align}
A(r,\varphi) &  =(-r^{2})\times\nonumber\\
&  [\frac{\cos\varphi}{32R^{3}}(32R^{2}-3M^{2}\{20+3R^{2}\gamma^{2}%
(10+(\pi-2\varphi)^{2}\}\nonumber\\
&  +32MR(9M\gamma-2)\sin\varphi)-6M\{3M\cos3\varphi-8MR\gamma(\pi
-2\varphi)\cos2\varphi\nonumber\\
&  +(10M-4R^{2}\gamma+9MR^{2}\gamma^{2})(\pi-2\varphi)\sin\varphi\}].
\end{align}
Eq.(12) then yields [1]%
\begin{equation}
\cos\psi=\frac{\left\vert A\right\vert }{[A^{2}+B(r)r^{2}]^{\frac{1}{2}}}.
\end{equation}
In a more convenient form, the final Rindler-Ishak expression for $\psi$ to be
used is [see (iii) above]%
\begin{equation}
\tan\psi=\frac{B^{1/2}r}{\left\vert A\right\vert }.
\end{equation}
The bending angle is defined by $\epsilon=\psi-\varphi$. The basic ingredients
are Eqs.(1), (11), (13) to be plugged into Eq.(15). We shall now discuss
specific cases.

\textbf{IV. Specific cases}

\textbf{Case 1}: $\varphi=0$, $M\neq0$, $k\neq0$ [This is Rindler-Ishak
choice; see also (ii)]. We get from the orbit Eq.(11)%
\begin{align}
r &  =16R^{3}/[4MR(8-3\pi R\gamma+3R^{2}\gamma^{2})-8R^{3}\gamma\nonumber\\
&  +3M^{2}\{5\pi(2+3R^{2}\gamma^{2})-4R\gamma(16+3R^{2}\gamma^{2})\}].
\end{align}
The SdS value of $r$ in the first order (in $M$) solution in Ref.[1] is
exactly $\frac{R^{2}}{2M}$, which can be recovered by neglecting the $M^{2}$
term in Eq.(16) and setting $\gamma=0$. For the unbound orbits associated with
lensing the distance of closest approach of a light wave to a galaxy will be
further from the center of the galaxy than the matter orbiting in it.
Consequently, for practical purposes for lensing one should consider the halo
regime where $\gamma R>\frac{M}{R}$ holds. We have been considering expansion
(11) in the Schwarzschild weak field regime due to $M$ so that $\frac{M}{R}%
<1$. We also assume $\gamma R<1$ so that $(\gamma R)^{2}$ terms can be neglected.

We shall implement these approximations on the results derived from the orbit
Eq.(11). Let us identify the leading order term in $r$ taking into account
practical data, say, for a typical galactic cluster Abell 2744, for which the
accurately observed central mass and Einstein radius respectively are
$M=2.90\times10^{18}$cm and $R_{\text{E}}=2.97\times10^{23}$cm [2]. The
function $r$ in Eq.(16) expands term by term to first order in $\gamma$ as
\begin{align}
r  &  =\frac{R^{2}}{2M}+\frac{R^{4}\gamma}{8M^{2}}-\frac{15\pi R}{32}%
-\frac{3\pi R^{3}\gamma}{64M}+3R^{2}\gamma-\frac{45\pi^{2}R^{2}\gamma}%
{2048}\nonumber\\
&  +\frac{225\pi^{2}M}{512}-\frac{45\pi MR\gamma}{8}+\frac{675\pi^{3}MR\gamma
}{8192}+O(M^{2})
\end{align}
For light moving in the halo regime, $R\geq R_{\text{E}}$, say, $R=3\times
10^{23}$cm and with $\gamma=7\times10^{-28}$ cm$^{-1}$ [12], it can be
verified that the condition $\gamma R>\frac{M}{R}$ holds and that the leading
value of $r$ is $\left(  \frac{R^{4}\gamma}{8M^{2}}\right)  $. Putting this
value of into $B(r)$, with $k=0.43\times10^{-56}$cm$^{-2}$ [12], we get the
signature protection%
\begin{equation}
B\left(  \frac{R^{4}\gamma}{8M^{2}}\right)  >0.
\end{equation}

Putting the value of $r$ from Eq.(16) into Eq.(13), we get%
\begin{align}
A &  =8R^{3}[M^{2}(78-48\pi R\gamma+90R^{2}\gamma^{2}+9\pi^{2}R^{2}\gamma
^{2})-32R^{2}]/[4MR(8-3\pi R\gamma\nonumber\\
&  +3R^{2}\gamma^{2})-8R^{3}\gamma+3M^{2}\{5\pi(2+3R^{2}\gamma^{2}%
)-4R\gamma(16+3R^{2}\gamma^{2})\}]^{2}.
\end{align}
Ignoring $M^{2}$ terms and putting $\gamma=0$ above, we recover the value of
$\left\vert A\right\vert $ in the SdS spacetime. The one sided bending angle
is given by $\epsilon=\psi-\varphi$ and let us calculate $\epsilon=\psi
=\psi_{0}$ when $\varphi=0$. Putting in Eq.(15) the values for $B(r)$ from
Eq.(1), $r$ from Eq.(16) and $A$ from Eq.(19), we get
\begin{align}
\tan\psi &  =\left(  \sqrt{1-\frac{2M}{r}+\gamma r-kr^{2}}\right)
\times\nonumber\\
&  2[4MR(8-3\pi R\gamma+3R^{2}\gamma^{2})-8R^{3}\gamma+3M^{2}\{5\pi
(2+3R^{2}\gamma^{2})\nonumber\\
&  -4R\gamma(16+3R^{2}\gamma^{2})\}]/[M^{2}(78-48\pi R\gamma+90R^{2}\gamma
^{2}+9\pi^{2}R^{2}\gamma^{2})-32R^{2}].
\end{align}
This is the exact formula for light deflection but is rather unilluminating.
Therefore, successively expanding in the first order of $\gamma$, and in the
second order in $M$, we find, for a small angle $\psi_{0}$ (or, $\tan\psi
_{0}\simeq\psi_{0}$), the following expression (see Appendix A for details,
see also Appendix C):
\begin{equation}
\psi_{0}\simeq\frac{2M}{R}\left[  1+\frac{15\pi M}{16R}-\frac{kR^{4}}{8M^{2}%
}\right]  -\gamma\left[  \frac{R}{2}+\frac{3\pi M}{4}+\frac{455M^{2}}%
{32R}+\frac{4M^{2}}{kR^{3}}\right]
\end{equation}
where we have retained only the leading order terms in the coefficient of
$\frac{2M}{R}$ and $\gamma$ assuming $\frac{M}{R}<1$. We find that all terms
in the last third bracket are positive, meaning that the effect of $\gamma$ is
to diminish the Schwarzschild bending even up to second order in $M$. This is
a core result of this paper.

We find that Eq.(21) reproduces the exact term $-\frac{\gamma R}{2}$ obtained
earlier in the literature by Edery and Paranjape [13] using Weinberg's method.
Thus, it is clear that the Rindler-Ishak method is not only correct but also
tells us more in the form of other \textit{new} terms showing, in the last
bracket, a clear interplay of $\gamma$, $M$ and $k$, each bearing its own
physical meaning. When the light orbits in the halo region, $R>R_{\text{E}}$,
it was already computed that the $\gamma-$ term dominates over the
Schwarzschild term: $\frac{\gamma R}{2}>\frac{2M}{R}$, i.e., the parameter
$\gamma$ plays the main role in the halo. At $R>R_{\text{E}}$, the effect of
$M\neq0$ would be practically negligible so that only repulsive bending
$-\frac{\gamma R}{2}$ will occur. However, the $(\gamma/k)-$ dependent term
could be comparable to the pure $\gamma-$ term in spite of the fact that in
the halo region, $\frac{M}{R_{E}}<<1$. We need realistic values to see that
fact. For the values stated around Eq.(17), we get: $\frac{\gamma R_{E}}%
{2}\simeq10^{-4}$, $\frac{4\gamma M^{2}}{kR_{E}^{3}}\simeq10^{-4}$ which shows
that the pure $\gamma-$ effect is of comparable magnitude with the combined
effect of $\gamma$, $M$ and $k$. The other terms are: $\frac{2M}{R_{E}}%
\simeq10^{-5}$ and $\frac{kR_{E}^{3}}{4M}\simeq10^{-7}$and are at least an
order of magnitude less than $\frac{\gamma R_{E}}{2}$. So, the Schwarzschild
bending $\frac{2M}{R_{E}}$ seems to be dwarfed by the $\gamma-$ terms.
However, as we see, this conclusion is sensitive to the exact magnitude of
$\gamma$. The problem is that, while the values of $M$, $R_{\text{E}}$ and $k$
are observationally known with adequate accuracy [2], the exact sign and value
of $\gamma$ are not yet conclusively known.

\textbf{Case 2:} $\varphi\neq0$, $M=0$, $k\neq0$. The exact solution of the
null geodesic is $\frac{1}{r}=\frac{\sin\varphi}{R}-\frac{\gamma}{2}$, which
for small angle $\varphi$ reduces to
\begin{equation}
\frac{1}{r}=\frac{\varphi}{R}-\frac{\gamma}{2}.
\end{equation}
Then the angle $\varphi$ corresponding to $r=r_{\text{max}}$ is
\begin{equation}
\varphi(r_{\text{max}})=\frac{\gamma R}{2}+\frac{kR}{\gamma}.
\end{equation}
According to the present method [see (iii)], we have, differentiating (22),
\begin{equation}
A(r_{\text{max}})=(-r_{\text{max}}^{2})\frac{1}{R}=-\frac{\gamma^{2}}{Rk^{2}}.
\end{equation}
Putting the values of $r_{\text{max}}=\frac{\gamma}{k}$, $\left\vert
A(r_{\text{max}})\right\vert $ and $B=B(r_{\text{max}})=1$ in (15)\ for small
$\psi$, we obtain,
\begin{equation}
\psi(r_{\text{max}})=\frac{kR}{\gamma}.
\end{equation}
The deflection is given by
\begin{equation}
\epsilon=\psi(r_{\text{max}})-\varphi(r_{\text{max}})=-\frac{\gamma R}{2}.
\end{equation}
The final result (26) is independent of $k$. Because of this independence, the
deflection in the pure conformal gravity with metric potential $B(r)=1+\gamma
r$ also gives the same deflection, viz., $\epsilon=-\frac{\gamma R}{2}$ (See
Appendix B for an alternative derivation).

\textbf{Case 3: }$\varphi=\pi/4$ [Rindler-Ishak choice], $M\neq0$, $k\neq0$.
Below we make some corrections. The Rindler-Ishak formula gives the bending
angle $\epsilon$ as%
\begin{equation}
\epsilon=\psi-\varphi\approx\tan(\psi-\varphi)=\frac{\tan\psi-\tan\varphi
}{1+\tan\psi\tan\varphi}=\frac{\tan\psi-1}{1+\tan\psi}.
\end{equation}
Proceeding in the similar manner as above, we can find the value of $\tan\psi$
and thence of $\epsilon$. Surprisingly the coefficient of $\gamma R$ becomes
identically zero. We do not give detailed expressions here but only state the
final result to leading order:%
\begin{equation}
\epsilon=\frac{\sqrt{2}M}{R}-\frac{3\pi\gamma M}{8}-\frac{\gamma^{2}R^{2}}%
{8}-\frac{k\gamma^{2}R^{4}}{4}-\frac{kR^{2}}{2}.
\end{equation}
Once again the deflection due to $\gamma$ is negative. Incidentally, the first
term is slightly different from that of Rindler and Ishak \textit{[their
Eq.(19)]}. Also, for $\gamma=0$, their expression \textit{[Eq.(18)]} for
$\tan\psi$ slightly modifies to%
\begin{equation}
\tan\psi=1+\frac{2\sqrt{2}M}{R}-kR^{2}.
\end{equation}
These modifications by no means alter their demonstration of the repulsive
effect of $k$ (in fact it is exactly the same). Even the pattern (loosely
speaking) of \textquotedblleft division by $2$\textquotedblright\ in the last
two pieces of $\tan\psi$ results in the corresponding pieces in $\epsilon$,
just as it is in Ref.[1]. The minor changes in the coefficient of $M$ have
come about because we have used the path equation at $\varphi=\pi/4$ to obtain
$r=\sqrt{2}R-3M+O(M^{2})$ rather than $r=\sqrt{2}R$.

\textbf{Case 4: }$\varphi=\pi/4$, $M=0$, $k\neq0$. Then%
\begin{equation}
r=\frac{2R}{\sqrt{2}-\gamma R}\text{, }A=\frac{2\sqrt{2}R}{(\sqrt{2}-\gamma
R)^{2}}\text{, }\tan\varphi=1\text{, }\tan\psi=\left(  \frac{\gamma^{2}%
R^{2}+4kR^{2}-2}{4\sqrt{2}\gamma R-2\gamma^{2}R^{2}-4}\right)  ^{1/2}.
\end{equation}
For small deflections, expanding successively in powers of $\gamma$ and first
power of $k$, we get%
\begin{equation}
\epsilon=\frac{\tan\psi-\tan\varphi}{1+\tan\psi\tan\varphi}\simeq-\frac
{\gamma^{2}R^{2}}{8}-\frac{k\gamma^{2}R^{4}}{4}-\frac{kR^{2}}{2}.
\end{equation}
This result can also be obtained directly from Eq.(28) by putting $M=0$.

\textbf{V. Summary }

As shown in Eq.(3), in the MKdS gravity too, which is more general than the
SdS gravity, the constant $k$ cancels out of the light orbit equation even
though $\gamma\neq0$, the latter fact distinguishing the MKdS metric from the
SdS metric. We should also remember that the corresponding parent theories are
generically very different; one is fourth order and the other is the second
order gravity. Nevertheless, in view of similar cancellation in the two
metrics, we investigated the applicability of the Rindler-Ishak method taking
the MKdS solution as an example. This is a nice example because Weyl conformal
gravity accommodates the successes of Schwarzschild gravity and has been the
subject of active research for several years. We should re-emphasize that we
are exclusively concerned here with the efficacy of the Rindler-Ishak method
and not with the well discussed but inconclusive values of and other
difficulties associated with the $\gamma-$ term. 

We first derived the exact null geodesic equation in the MKdS gravity. Next,
we perturbatively solved the equation up to the order $M^{2\text{ }}$though
the zeroth order equation is different from that in standard Schwarzschild
gravity. Whatever follows below are the results obtained after the detailed
solution is plugged into the Rindler-Ishak procedure, which we have faithfully implemented.

We should note that, generally speaking, none of the quantities $k$, $\gamma$
and $M$ should be zero. (We can nonetheless set one or the other to zero as
limiting cases). Then Eq.(21) at once shows the influence of $k$ both in the
Schwarzschild and conformal sector. The equation nicely reproduces the
attractive Schwarzschild bending due to $M>0$ as well as the repulsive bending
due to cosmological constant $k>0$ and the Weyl conformal parameter $\gamma
>0$. Eq.(21) combines in one place deflections at various orders obtained by
Bodenner \& Will [15], by Rindler \& Ishak [1] and by Edery \& Paranjape [13].
Because of $M^{2}$ order, there appeared new terms in the conformal sector. In
particular, there is a second order correction $-\frac{3\pi\gamma M}{4}%
=-\frac{3\pi}{4}\times\gamma R\times\frac{M}{R}$ to the first order term
$-\frac{\gamma R}{2}$ obtained in Ref.[13], which also diminishes the
Schwarzschild one way bending.

There is absolutely no problem in the case $\varphi=\pi/4$, $k=0$, $M=0$
[Eq.(28)] because the expression for $\epsilon$ does not blow off. In this
case, the term $-\frac{\gamma R}{2}$ of course does not appear, but one
notices that the effect is $-\frac{\gamma^{2}R^{2}}{8}$, which remains
unaltered by the choice of sign for $\gamma$. In either case ($\varphi=0$ or
$\pi/4$), the conformal parameter $\gamma$ can always be set to zero at will,
but then one ends up with the already discussed SdS case. For the case $M=0$,
we determine the maximum allowed value of $r=r_{\text{max}}$ and
correspondingly determine $\epsilon=-\frac{\gamma R}{2}$ [Eq.(26)]. A direct
integration given in Appendix B also supports this result. Since $\epsilon$ is
independent of $k$, this then is also the deflection in the pure conformal
gravity defined by $M=0$, $k=0$ so that $B(r)=1+\gamma r$. Overall, the
conclusion is that the Rindler-Ishak method can be applied to more general
situation than SdS and that it leads to the same result up to second order as
obtained by conventional perturbative method (See Appendix C).

The implication of the last term in Eq.(21), viz., $-\frac{4\gamma M^{2}%
}{kR^{3}}$, is very interesting and seems to provide the background for a
certain postulate. To have some idea how, let us estimate the terms in Eq.(21)
for a light ray grazing a stellar sized massive object, say, the Sun itself,
and that $\gamma\approx10^{-28}$cm$^{-1}$, $k\approx10^{-56}$cm$^{-2}$, the
values being already independently estimated. We then have the following
numerology:%
\begin{equation}
M_{\odot}=1.48\times10^{5}\text{ cm, }R_{\odot}=6.96\times10^{10}\text{ cm, }%
\end{equation}
so that
\begin{equation}
\frac{2M_{\odot}}{R_{\odot}}=4.24\times10^{-6}\text{, }\frac{30\pi M_{\odot
}^{2}}{16R_{\odot}^{2}}=2.65\times10^{-11}\text{, }-\frac{kR_{\odot}^{4}%
}{4M_{\odot}}\approx-10^{-19}%
\end{equation}%
\begin{equation}
-\frac{\gamma R_{\odot}}{2}\approx-10^{-18}\text{,}-\frac{3\pi\gamma M_{\odot
}}{4}\approx-10^{-23}\text{,}-\frac{4\gamma M_{\odot}^{2}}{kR_{\odot}^{3}%
}\approx-10^{6}.
\end{equation}
We find that, while all the other terms above are quite small compared to
$\frac{2M_{\odot}}{R_{\odot}}$, the spoiling\textit{ }term is the very last
one. If we had limited ourselves to the direct integration of the null
trajectory in which $k$ does not appear, we would have missed this term. This
term dominates at the limb of the Sun, giving rise to a total bending,
$\psi_{0}\approx$ $-10^{6}$, which is negative and means repulsion $10^{12}$
times bigger than the first order Schwarzschild attraction! Certainly this is
contrary to our experience. Moreover, with $M=M_{\odot}$, for any $R\geq
R_{\odot}$, one finds that the values of $\psi_{0}$ continue to remain only
negative. Without that spoiling term, however, the effect of $\gamma$ would
indeed be negligible near the Sun so that $\psi_{0}$ would not appreciably
differ from the positive Schwarzschild value, which does not appear to be the
case here.

One possible viewpoint is to postulate\textit{ }that, at the solar scale, MKdS
is predominantly only SdS [negligible $\gamma-$effect like in the first two
terms in (34)], while its \textit{genuine} applicability lies at the galactic
scale. In fact, the deflection on the galactic cluster scale tells a quite
different story: Consider again Abell 2744 [2], with $\frac{\gamma}{k}%
\approx10^{28}$cm, the term $-\frac{4\gamma M^{2}}{kR_{E}^{3}}\approx-10^{-4}%
$, which is comparable in order of magnitude with that of $-\frac{\gamma R}%
{2}$ and is at most an order of magnitude higher than the attractive
Schwarzschild term if we believe in the value of $\gamma$ used here. In any
case, the overall repulsion thus obtained in the halo can always be converted
to the desired attractive bending if the numerical value of $\gamma$ is
slightly altered for which possibilities certainly exist. The main conclusion
is that, with $\gamma\sim0$ in the solar system, the huge repulsion
($-\frac{4\gamma M_{\odot}^{2}}{kR_{\odot}^{3}}\approx-10^{6}$) term can be
avoided. This implies that $\gamma$ could be physically relevant \textit{only}
on the galactic cluster scale and not on the solar scale. The fact that
$\gamma$ operates in the galactic halo has been conjectured in the literature,
but here we find its support from a completely different viewpoint, viz., from
the Rindler-Ishak bending.

\textbf{Acknowledgment }

One of us (AB) wishes to thank the authorities of the Universit\`{a} La
Sapienza, Roma, for financial support during the project work. KKN and RI wish
to thank Guzel N. Kutdusova for technical assistance.

\textbf{References}

[1] W. Rindler and M. Ishak, Phys. Rev. D \textbf{76}, 043006 (2007).

[2] M. Ishak, W. Rindler, J. Dossett, J. Moldenhauer and C. Allen, Mon. Not.
Roy. Astron. Soc. \textbf{388}, 1279 (2008).

[3] M. Sereno, Phys.Rev. D \textbf{77}, 043004 (2008); Phys. Rev. Lett.
\textbf{102}, 021301 (2009).

[4] R. Kantowski, B. Chen and X. Dai (arXiv:0909.3308), to appear in ApJ.

[5] T. Sch\"{u}cker, General Relativity and Gravitation, \textbf{41}, 1595
(2009); \textit{ibid}, \textbf{41}, 67 (2009), see also: arXiv:1006.3234.

[6] F. Simpson, J. A. Peacock and A. F. Heavens, Mon. Not. Roy. Astron.
Soc.\textbf{ 402}, 2016 (2009).

[7] M. Ishak, W. Rindler and J. Dossett, Mon. Not. Roy. Astron. Soc.
\textbf{403}, 2152 (2010).

[8] M. Park, Phys. Rev. D \textbf{78,} 023014 (2008).

[9] I. Khriplovich and A. Pomeransky, Int. J. Mod. Phys. D\textbf{17}, 2255 (2008).

[10] Amrita Bhattacharya, Guzel M. Garipova, Alexander A. Potapov, Arunava
Bhadra and Kamal K. Nandi (arXiv:1002.2601).

[11] M. Ishak and W. Rindler (arXiv:1006.0014), and references therein, to
appear in GRG.

[12] P. D. Mannheim and D. Kazanas, Astrophys. J. \textbf{342}, 635 (1989).

[13] A. Edery and M. B. Paranjape, Phys. Rev. D \textbf{58}, 024011 (1998).

[14] S. Pireaux, Class. Quant. Grav. \textbf{2}1, 4317 (2004).

[15] J. Bodenner and C.M. Will, Am. J. Phys. \textbf{71}, 770 (2003).

[16] S. Weinberg, 1972, \textit{Gravitation \& Cosmology }(John Wiley \& Sons,
New York), p.189.

\begin{center}
\textbf{Appendix A}
\end{center}

For the benefit of the readers we give below the exact steps leading to the
expression (21). After putting the value of $r$ from Eq.(16) into Eq.(20) and
simplifying, we obtain%
\begin{equation}
\tan\psi_{0}=\frac{1}{S}\left[
\begin{array}
[c]{c}%
2(P+Q)\left\{  1+\gamma M-\frac{M^{2}(8-3\pi\gamma R+3\gamma^{2}R^{2})}%
{2R^{2}}\right. \\
\left.  -\frac{Q}{8R^{3}}+\frac{16\gamma R^{3}}{P+Q}-\frac{256kR^{6}%
}{(P+Q)^{2}}\right\}  ^{1/2}%
\end{array}
\right]  \tag{A1}%
\end{equation}
where%
\begin{equation}
P\equiv4MR(8-3\pi\gamma R+3\gamma^{2}R^{2})-8\gamma R^{3} \tag{A2}%
\end{equation}%
\begin{equation}
Q\equiv3M^{2}\{5\pi(2+3\gamma^{2}R^{2})-4\gamma R(16+3\gamma^{2}R^{2})\}
\tag{A3}%
\end{equation}%
\begin{equation}
S\equiv M^{2}(78-48\pi R\gamma+90R^{2}\gamma^{2}+9\pi^{2}R^{2}\gamma
^{2})-32R^{2}. \tag{A4}%
\end{equation}
In order to extract the contribution due purely to $\gamma$, we expand the
right hand side in the first power of $\gamma$, and get%
\begin{align}
\tan\psi_{0}  &  \simeq\frac{2TU}{78M^{2}-32R^{2}}\nonumber\\
&  +\frac{2\gamma T}{V}\left[  M+\frac{3\pi M^{2}}{2R}+\frac{24M^{3}}{R^{2}%
}+\frac{16R^{3}}{T}-\frac{265kR^{7}W}{M^{3}(15\pi M+16R)^{3}}\right]  \tag{A5}%
\end{align}
where%
\[
T\equiv30\pi M^{2}+32MR
\]%
\begin{equation}
U\equiv\left(  1-\frac{4M^{2}}{R^{2}}-\frac{15\pi M^{3}}{4R^{3}}%
-\frac{256kR^{6}}{T^{2}}\right)  ^{1/2} \tag{A6}%
\end{equation}%
\begin{equation}
V\equiv U\left[  156M^{2}-64R^{2}-\frac{2RW}{39M^{2}-16R^{2}}+\frac{24\pi
RM^{3}(15\pi M+16R)}{(39M^{2}-16R^{2})^{2}}\right]  \tag{A7}%
\end{equation}

\begin{equation}
W\equiv48M^{2}+3\pi MR+2R^{2}. \tag{A8}%
\end{equation}
We first expand the second term in (A5) in powers of $M$ obtaining%
\begin{equation}
\psi_{0}^{\text{MKdS}}\simeq-\gamma\left[  \frac{R}{2}+\frac{3\pi M}{4}%
+\frac{455M^{2}}{32R}+\frac{4M^{2}}{kR^{3}}\right]  . \tag{A9}%
\end{equation}
Next let us put $\gamma=0$ in (A5) in order to obtain pure Schwarzschild terms
from%
\begin{equation}
\tan\psi_{0}^{\text{SdS}}\simeq\frac{2TU}{78M^{2}-32R^{2}}. \tag{A10}%
\end{equation}
The deflection is already known to be positive. To ensure it, Rindler and
Ishak in their SdS treatment prescribed $\left\vert A\right\vert $ instead of
just $A$. This prescription is the same as changing the denominator of (A10)
into $32R^{2}-78M^{2}$ because $R>>M$ and $TU>0$. For small $\psi_{0}$, and
with $U\simeq1-\frac{2M^{2}}{R^{2}}-\frac{15\pi M^{3}}{8R^{3}}-\frac
{128kR^{6}}{T^{2}}$, the first term in (A5) then results in%
\begin{equation}
\psi_{0}^{\text{SdS}}=\frac{2(30\pi M^{2}+32MR)\left[  1-\frac{2M^{2}}{R^{2}%
}-\frac{15\pi M^{3}}{8R^{3}}-\frac{128kR^{6}}{T^{2}}\right]  }{32R^{2}%
-78M^{2}}, \tag{A11}%
\end{equation}
which, when expanded in powers of $M$, yields
\begin{equation}
\psi_{0}^{\text{SdS}}\simeq\frac{2M}{R}\left[  1+\frac{15\pi M}{16R}%
-\frac{kR^{4}}{8M^{2}}\right]  . \tag{A12}%
\end{equation}
The total one way deflection is of course%
\begin{equation}
\psi_{0}=\psi_{0}^{\text{SdS}}+\psi_{0}^{\text{MKdS}} \tag{A13}%
\end{equation}
which is just the Eq.(21) in the text.

\begin{center}
\textbf{Appendix B}
\end{center}

Consider the null geodesic equation when $M=0$:
\begin{equation}
\frac{d^{2}u}{d\varphi^{2}}=-u-\frac{\gamma}{2} \tag{B1}%
\end{equation}
which has an exact solution%
\begin{equation}
u=\frac{1}{r}=\frac{\sin\varphi}{R}-\frac{\gamma}{2}. \tag{B2}%
\end{equation}
The metric for $M=0$ is
\begin{equation}
B(r)=1+\gamma r-kr^{2}. \tag{B3}%
\end{equation}
Weinberg's method allows a direct integration giving the deflection as%
\begin{align}
\Delta\varphi &  =\int_{R}^{\infty}\left[  \frac{r^{4}(1+\gamma R)}{R^{2}%
}-r^{2}-\gamma r^{3}\right]  ^{-1/2}dr\nonumber\\
&  \simeq\int_{R}^{\infty}\left(  \frac{r^{4}}{R^{2}}-r^{2}\right)
^{-1/2}dr-\frac{\gamma R}{2}\int_{R}^{\infty}\left(  \frac{r^{4}}{R^{2}}%
-r^{2}\right)  ^{-1/2}\frac{rdr}{R+r}. \tag{B4}%
\end{align}
The last line is obtained after expanding the integrand in small $\gamma R$.
As noted by Edery and Paranjape [13], $k$ has cancelled out of the integrand.
It is to be expected since the path equation does not contain $k$ either.
Therefore, even for pure conformal gravity, when $B(r)=1+\gamma r$, the above
result holds true. The integration in (B4) yields%
\begin{equation}
\Delta\varphi-\frac{\pi}{2}=-\frac{\gamma R}{2}. \tag{B5}%
\end{equation}
This result supports the fact that pure conformal gravity is repulsive. The
factor $\frac{\pi}{2}$ appears due to the conventional definition of the
bending angle. This factor does not appear according to Rindler-Ishak
definition [1] of the azimuthal angle, but the final result is always the same
once it is corrected for, which is indeed the case as shown in Eq.(26) in the text.

\begin{center}

\textbf{Appendix C}
\end{center}

It is well known that the structure of complete solution of an ordinary
differential solution has two parts: The Characteristic Function (CF) is one
part when the right hand side is zero and the Particular Integral (PI) is the
other part when the same is nonzero. One might argue that terms proportional
to $\sin\varphi$, $\cos\varphi$ of the perturbative differential equation (8)
[by the same token, also of Eq.(7)] are trivial CFs and hence should be
discarded from the nontrivial PI part of the solutions. Indeed, in the SdS
light trajectory (with $\gamma=0$), the full solution of the perturbed zeroth
order equation $\frac{d^{2}u_{0}}{d\varphi^{2}}+u_{0}=0$ contains only
$u_{0}\varpropto\sin\varphi$, $\cos\varphi$ as CFs and no PI because the right
hand side is identically zero. So one might wish to avoid their repetitive
occurrence in the other two higher order perturbed solutions. This can be
easily achieved by choosing the constants of integration. However, in the case
$\gamma\neq0$, the full zeroth order solution is%
\begin{equation}
u_{0}=\text{CF+PI}=(C_{1}\sin\varphi+C_{2}\cos\varphi)+\left(  \frac
{\cos\varphi}{R}-\frac{\gamma}{2}\right)  . \tag{C1}%
\end{equation}
We can choose $C_{1}=0$ so that the reduced CF is $C_{2}\cos\varphi$. If we do
not want its repetition in the nontrivial PI part $\left(  \frac{\cos\varphi
}{R}-\frac{\gamma}{2}\right)  $, there are two ways: We can simply delete
$\frac{\cos\varphi}{R}$ from it and choose $C_{2}=1/R$ assuming $R$ to be the
distance of closest approach. Alternatively, we can set all the arbitrary CF
constants to zero and retain only the PI \textit{as it is}, even if it
includes\textit{ }the trivial term. In either way, we would end up with the
same $u_{0}=$ $\frac{\cos\varphi}{R}-\frac{\gamma}{2}$. We want to examine the
conventional Bodenner-Will method [15] for the MKdS solution using both the
ways (viz., with or without CFs in PI).

The structure of the exact full solution (CF+PI) of the three second order
perturbative equations (4), (7) and (8) would involve six arbitrary
integration constants $C_{1}$, $C_{2}$, $C_{3}$, $C_{4}$, $C_{5}$ and $C_{6}$
so that in the Schwarzschild case, we find%
\begin{align}
u  &  =[C_{1}\sin\varphi+C_{2}\cos\varphi\nonumber\\
&  +(C_{3}+C_{4})\sin\varphi+(C_{5}+C_{6})\cos\varphi]\nonumber\\
&  +\frac{15M^{2}}{8R^{3}}\cos\varphi+\frac{M}{2R^{2}}\left[  3-\cos
2\varphi\right]  +\frac{3M^{2}}{16R^{3}}[20\varphi\sin\varphi+\cos3\varphi].
\tag{C2}%
\end{align}
The form of the solution considered by Bodenner and Will [9] is equivalent to
choosing
\begin{equation}
C_{1}=0\text{, }C_{2}=\frac{1}{R}\text{, }C_{3}+C_{4}=0\text{, }C_{4}%
+C_{6}=-\frac{15M^{2}}{8R^{3}}, \tag{C3}%
\end{equation}
which removes the trivial CF $\frac{15M^{2}}{8R^{3}}\cos\varphi$ from the PI
part [last line in (C2)]. Once the constants are chosen, the solution is
uniquely fixed. The analysis of the solution $u$ with the choice (C3) is well
known and need not be discussed here. We could as well choose $C_{1}=1/R$,
$C_{2}=0$ in the zeroth order solution, and correspondingly the PIs would
change, from which we can remove the CF in a similar manner as indicated above.

As promised, let us consider the MKdS solution, first setting the trivial CFs
to zero but \textit{including} them in the PIs as they naturally appear. Then
we get%
\begin{equation}
u=u_{0}+u_{1}+u_{2} \tag{C4}%
\end{equation}
where $u_{0}$, $u_{1}$ and $u_{2}$ are given by Eqs.(5), (9) and (10)
respectively. For $\gamma\neq0$, this form of the solution is obviously
different from that considered by Bodenner and Will because the last two PIs
($u_{1}$ and $u_{2}$) \textit{do} contain terms proportional to trivial
$\cos\varphi$. [When $\gamma=0$, $u$ consists of last three terms in (C2)].
The Bodenner-Will method has been applied to the asymptotically flat spacetime
allowing the standard radial coordinate $r\rightarrow\infty$. So they take
$u\rightarrow0$ at $\varphi=\pi/2+\delta$ and for small $\delta$, $\sin
\delta\simeq\delta$, $\cos\delta\simeq1$. But $u\rightarrow0$ is not allowed
in the present non-flat metric; instead one should take\footnote{The
modification to $r_{\text{max}}$ of Case 2, Sec.IV by the presence of $M$ is
problematic, because the full equation $B(r)=0$ yields unwieldy roots. Hence
we simply assume that there \textit{exists} some $r=r_{\text{max}}$ or
$u=u_{\text{min}}\sim0$.} $u=u_{\text{min}}.$ Let us see what the
Bodenner-Will method yields for the deflection angle.

Ignoring the small terms proportional to $\delta^{2}$, ($R\gamma$)$^{2}$ and
($R\gamma$)$^{3}$ in the equation resulting from (C4), we finally obtain%
\begin{align}
u_{\text{min}}  &  =\frac{1}{16R^{3}}\left[  4MR(8-3\pi R\gamma)-8R^{2}%
(R\gamma+2\delta)\right. \nonumber\\
&  \left.  -3M^{2}\left\{  64R\gamma-13\delta+2\pi(4R\gamma\delta-5)\right\}
\right]  . \tag{C5}%
\end{align}
This yields the value for $\delta$ as%
\begin{equation}
\delta=\frac{2(15M^{2}\pi+16MR-96M^{2}R\gamma-6M\pi R^{2}\gamma-4R^{3}%
\gamma)-16R^{3}u_{\text{min}}}{16R^{2}-69M^{2}+24M^{2}\pi R\gamma}. \tag{C6}%
\end{equation}
We straightforwardly expand $\delta$ to find that
\begin{align}
\delta &  \simeq\frac{2M}{R}+\frac{15\pi M^{2}}{8R^{2}}-\gamma\left[  \frac
{R}{2}+\frac{3\pi M}{4}+\frac{423M^{2}}{32R}\right] \nonumber\\
&  +u_{\text{min}}\left\{  \left(  \frac{3\pi\gamma}{2}-\frac{39}{16R}\right)
M^{2}-R\right\}  . \tag{C7}%
\end{align}

Next, if we had \textit{excluded} the CFs from PI (\textit{\`{a} la} Bodenner
and Will [15]), by suitably choosing constants we would obtain, under the same
approximation,
\begin{align}
u_{\text{min}} &  =\frac{1}{16R^{3}}\left[  4MR(8-3\pi R\gamma)-8R^{2}%
(R\gamma+2\delta)\right.  \nonumber\\
&  \left.  -3M^{2}\left\{  64R\gamma-23\delta+2\pi(4R\gamma\delta-5)\right\}
\right]  ,\tag{C8}%
\end{align}
which would yield%
\begin{align}
\delta &  \simeq\frac{2M}{R}+\frac{15\pi M^{2}}{8R^{2}}-\gamma\left[  \frac
{R}{2}+\frac{3\pi M}{4}+\frac{453M^{2}}{32R}\right]  \nonumber\\
&  +u_{\text{min}}\left\{  \left(  \frac{3\pi\gamma}{2}-\frac{69}{16R}\right)
M^{2}-R\right\}  \tag{C9}%
\end{align}
So in (C9), we find $-\frac{453\gamma M^{2}}{32R}$ instead of $-\frac
{423\gamma M^{2}}{32R}$, and $-\frac{69M^{2}u_{\text{min}}}{16R}$ instead of
$-\frac{39M^{2}u_{\text{min}}}{16R}$, while other terms remain the same.
That's all there is to it. However, note that differences occur only in the
\textit{third} order of smallness, $\frac{\gamma M^{2}}{R}=\gamma R\times
\frac{M^{2}}{R^{2}},$ which should not concern us as our purpose was to
calculate effects only up to second order using the orbit equation to that
order. Genuine third order effects would require the orbit equation in the
third order in $M$. So we could as well delete those third order terms, none
of which is genuine, from Eq.(21), (C7) and (C9) but we have nevertheless
displayed them only to demarcate where the series ends. The remaining terms
factored with $u_{\text{min}}$ are computed in the footnote.\footnote{The
exact value of $r_{\text{max}}$ (or $u_{\text{min}}$) from full $B(r)=0$ is
rather messy. Hence, to simplify matters, let us \textit{assume} that
$u_{\text{min}}=\frac{k}{\gamma}$ and that $0<R<<\frac{\gamma}{k}\Rightarrow$
$kR<<\gamma$. With these assumptions, the first two terms factored with
$u_{\text{min}}$ are at least of the third order of smallness, hence
ignorable. The last one is $\frac{kR}{\gamma}<<1$ but its order of smallness
depends on how small $R$ is compared to $\frac{\gamma}{k}$. An order of
magnitude calculation assuming $\frac{\gamma}{k}\sim10^{28}$cm is in place. In
the light-grazing-sun bending, numerology suggests that $\frac{kR_{\odot}%
}{\gamma}\sim\left(  \frac{M_{\odot}}{R_{\odot}}\right)  ^{3}$, meaning third
order of smallness, hence again ignorable. But for galactic clusters, say,
Abell 2744, the values suggest that, while the first two terms are truly of
the third order of smallness, the term $-\frac{kR_{E}}{\gamma}\sim-10^{-5}$ is
in the first order. It is remarkable that the same first order effect follows
also from the Rindler-Ishak bending.} The results of this Appendix therefore
clearly demonstrate that differences in solution by inclusion or exclusion of
trivial terms in PI (or choices of zeroth order $\sin\varphi$, $\cos\varphi$)
lead to no differences in the final result for observable bending within the
considered order.

As stated in Sec.III, we wanted to faithfully adhere to the Rindler-Ishak [1]
form of the solution (which they gave for $\gamma=0$). Also, we recover\ from
Eq.(11) the same form (\textit{sans} $10\sin\varphi$), now up to second order
in $M$, as derived in Ishak \textit{et al} [2]. They chose the definition of
azimuthal angle (differing by $\frac{\pi}{2}$ from the conventional one) that
we also have maintained for the purpose of easy comparison with their
expressions for $\frac{1}{r}$. Certainly, the final result for bending does
not depend on this definition as shown in this Appendix, where the choice of
the azimuthal angle has been conventional, as in Bodenner and Will [15]. The
remarkable similarity of the terms up to second order in (C7) and (C9) above
with those in Eq.(21) shows that the bending derived via Rindler-Ishak method
is indeed correct.

\end{document}